\documentstyle[12pt]{article}
\begin{document}
\begin{titlepage}
\begin{flushright}
     \hfill \\
\end{flushright}
\vspace{1.0 in}
\def\as{$\alpha_s$}
\centerline{\bf{Charm Production at RHIC to
$O(\alpha_{s}^{3})$}}
\vskip 0.2true in
\centerline{Ina Sarcevic and Peter Valerio}
\centerline{Department of Physics}
\centerline{University of Arizona}
\centerline{Tucson, AZ 85721}
\vskip 1.3true in
\begin{abstract}
We present results on rapidity and transverse momentum
distributions of inclusive charm quark production in
heavy-ion collisions at RHIC, including the next-to-leading
order, $O(\alpha_s^3)$, radiative corrections and the
nuclear shadowing effect.
We find the effective, nuclear K-factor
to be $K(y)\approx 1.4$ for $\mid y\mid \leq 3$ in the rapidity
distribution,
while
$1\leq K(p_{T}) \leq 3$ for
$1GeV\leq p_T \leq 6GeV$ in the $p_T$ distribution.
We incorporate multiple parton scatterings in our calculation of
the fraction of all
central events that contain at least one charm quark pair.
We obtain the
effective $A$-dependence of the charm cross sections.
Finally, we comment on the possibility
of detecting the quark-gluon plasma signal as an
enhanced charm production in heavy-ion collisions at RHIC.

\end{abstract}
\end{titlepage}
\vskip 0.05true in
\vfil\eject
\par
The main goal of the future heavy-ion colliders, such as
RHIC, is to
study the
properties of nuclear matter under extreme conditions, and
in particular to search for  the formation of  a
new state of matter, the
quark-gluon plasma [1].
With the assumption
that high energy nuclear collisions lead to the
thermalized system,
RHIC  energies are sufficiently
large to
produce very dense matter,
of the order of several $GeV/fm^{3}$,
well above the critical density necessary to create
the quark-gluon plasma
[2].  The problem of finding a clean, detectable signal
for this phase is
currently one of the most challenging theoretical
problems.
In the last few years, the possible signals, such as
thermal photons, dileptons and $J/\Psi$ suppression,
have been
investigated in detail and
found to have yields comparable in magnitude
with those expected from a simple
extrapolation of hadronic collisions [3].
Recently,  open charm production in heavy-ion collisions
has been proposed as an
elegant method for probing the possible formation of the
quark-gluon plasma [4].
However, in order
to determine
whether the enchanced charm production at RHIC can be
unambiguously interpreted as a signal of QGP,
one needs to have control of the other sources of
charm production.  The standard competing
process is  charm production through the hard collisions
of partons inside the nuclei.
In addition to being relevant as
a background for the signal of quark-gluon plasma, this
type of charm production is useful tool for studying
the perturbative aspects of strong interactions and for
determining
the nuclear screening/shadowing effect on the gluon
distribution in a nucleus.
\par
In this letter we present results of  calculations of the
rapidity and transverse momentum distributions of
inclusive charm quark production in
Au-Au collisions at RHIC, including the $O(\alpha_s^3)$
radiative
corrections and the
nuclear shadowing effect.
We determine the size of the
gluon contribution to the charm production at RHIC.
We also calculate the fraction of central
and inelastic events at RHIC that will contain charm quarks
which satisfy unitarity constraints by
properly taking into account multiple nucleon scatterings.
In our calculations we use the
latest set of two-loop evolved parton densities in a nucleon
obtained from
global fits of data
from deep inelastic lepton-nucleon collisions [5].
We take the nuclear
shadowing effect in the quark distribution from
the recent measurements in
deep inelastic lepton-nucleus collisions [6].
We assume that the
amount of
shadowing present in the gluon distribution is the same as
in the quark.
To illustrate the
importance of next-to-leading order contributions
and
the nuclear shadowing effects, we determine the {\it effective}
(i.e. nuclear)  K-factor
defined as the ratio of the particular distribution
to the leading-order distribution  without any
nuclear effects.
We show that this K-factor is very different than
the one in hadronic collisions,
and that in general, it can not be
approximated
by a constant.
We also determine the
effective A-dependence of the charm cross sections.
Finally, we
comment on our results for the
number of charm quarks produced in central
rapidity region in central
Au-Au collisions in the context of quark-gluon plasma
signatures at RHIC.

\par
In perturbative QCD,
the inclusive cross section for
charm production
in
nuclear collisions is obtained by
convolution of parton densities in nuclei with a hard scattering
parton cross
section [7].
In our calculation,
for the parton cross sections,
we include leading-order subprocesses, $O(\alpha_s^2)$, such as
$ q+\bar q\rightarrow Q+\bar Q$ and
$ g+g\rightarrow Q+\bar Q$,
and next-to-leading order contributions, $O(\alpha_s^3)$, such as
$ q+\bar q\rightarrow Q+\bar Q +g$,
$ g+q \rightarrow Q+\bar Q+g$,
$ g+\bar q \rightarrow Q+\bar Q+\bar q$ and
$ g+g \rightarrow Q+\bar Q+g $.  The double differential inclusive
distribution of
charm production in central
AA collisions can be written as
\begin{equation}
\frac{d N_c}{d^{2}p_{T}dy}=T_{AA}(0)
\frac{d{\sigma}_{c}
}{d^{2}p_{T}dy},
\end{equation}
where the double differential inclusive cross section
is given by
\begin{equation}
\frac{d{\sigma}_{c}
}{d^{2}p_{T}dy}
=
\sum_{i,j}^{partons}
\int dx_{a} dx_{b}
F_{i}^{A}(x_{a},Q^{2})F_{j}^{A}(x_{b},Q^{2})
\frac{d\hat{\sigma}_{i,j}
(Q^2, m_c, \hat s)}{d^{2}p_{T}dy},
\end{equation}
and $T_{AA}(0)$ is the nuclear overlapping density at zero impact
parameter,
$F_{i}^{A}(x,Q^2)$ is the parton structure function in a
nucleus,
$x_a$ and
$x_b$ are the fractional momenta of the incoming partons,
$\hat s$ is the parton-parton c.m. energy ($\hat s=x_a x_b s$).
The parton differential cross section calculated to
$O(\alpha_{s}^{3})$ can be written as
[8]
\begin{equation}
\frac{d\hat{\sigma}_{i,j}}{d^{2}p_{T}dy }=
\frac{\alpha^{2}_{s}}{\hat s}h^{(0)}_{i,j}+\frac{\alpha^{3}_{s}}
{2\pi \hat s^{2}}h^{(1)}_{i,j}.
\end{equation}
Expressions
for the functions $h^{(0)}_{i,j}$ and
$h^{(1)}_{i,j}$
can be found in Ref. 8.
The coupling constant $\alpha_s(Q^2)$ that appears in Eq. (3)
is given by
\begin{equation}
\alpha_s (Q^2)={12\pi\over
{(33-2N_f)\ln {Q^2\over{\Lambda^2}}}}[1-{6
(153-19N_f){\ln \ln {Q^2/
{\Lambda^2}}}\over{{(33-2N_f)}^2\ln {Q^2/\Lambda^2}}}],
\end{equation}
where $Q^2$ is the renormalization scale,
$\Lambda$ is the QCD scale
parameter and $N_f$ is the number of flavors.
We take the factorization scale in the structure functions to be
$2m_c$ and we consider the renormalization scales
$Q=m_c$ and $Q=2m_c$.  We do not evolve the structure
functions
below $Q=2m_c$, because for $Q^2\leq 8.5GeV^2$
their behavior is not very well known [9].
For the mass of the charm quark we use
$m_c=1.5GeV$.

The double differential distribution for
inclusive charm
quark production in
{\it inelastic } Au-Au collisions can be obtained from
Eq. (1) by replacing
$T_{AA}(0)$ with
$A^2/\sigma_{inelastic}^{AA}$, where $\sigma_{inelastic}^{AA}\approx
4\pi R_A^2$.

To obtain the number of nucleon-nucleon collisions per unit of
transverse
area at fixed impact parameter, we consider the
nuclear overlapping function [10]
\begin{equation}
T_{AA}(b)=\int d^{2}b_{1}
T_{A}(\mid\vec{b}_{1}\mid)
T_{A}(\mid\vec{b} - \vec{b}_{1}\mid),
\end{equation}
where
the nuclear profile function,
$T_A(b)$, is the
nuclear density integrated over the longitudinal size, i.e.
\begin{equation}
T_{A}(b)=
\int_{-\infty}^{\infty}dz \rho_{A}(\sqrt{b^{2}+z^{2}})
\end{equation}
For nuclear density we use the
Woods-Saxon
distribution [11]  given by
\begin{equation}
\rho(r)=\frac{n_{0}}{[1+e^{{(r-R_{A})}/d}]}.
\end{equation}
The density and the
nuclear overlapping function are normalized so that
$\int d^{3}  r \rho( r) = A$ and
$\int d^{2} b T_{AA}(b)=A^{2}$.
For central collisions
the
overlapping function
can be approximated
by $T_{AA}(0)={A}^{2}/\pi {R_{A}}^{2}$, which gives
$T_{Au-Au}(0)=
30.7 mb^{-1}$.
\par
The nuclear parton distribution, if
nucleons were independent, would be given
as A times the parton structure function in a nucleon.
However, at high
energies,
the parton densities become so large that the sea quarks and
gluons overlap
spatially and
the nucleus can not be viewed as a collection of uncorrelated
nucleons.
This happens when the longitudinal size of the parton, in
the infinite
momentum frame of the nucleus, becomes larger than the
size of the nucleon.  Partons from different nucleons start
to interact and through annihilation effectively reduce the parton
density in a nucleus.  When partons inside the nucleus completely
overlap, there reach a saturation point.  Motivated by this simple
parton picture
of the nuclear shadowing effect and taking into account
the $A^{1/3}$
dependence obtained by considering the modified, nonlinear
modifying factor to the
Altarelli-Parisi equations with gluon recombination included, the
parton structure function in a nucleus can be written as [12]
\begin{equation}
R(x,A)\equiv \frac {F_{i}^{A}(x,Q^{2})}{A F_{i}^{N}(x,Q^{2})}
=\left\{\begin{array}{ll}
1-\frac{3}{16}x+\frac{3}{80}&.2<x\leq 1\\
1&x_{n}<x\leq .2\\
1-D(A^{1/3}-1)
\frac{1/x-1/x_{n}}
{1/x_{A}-1/x_{n}}&x_{A}\leq x\leq x_{n}\\
1-D(A^{1/3}-1)&0<x<x_{A}
\end{array}
\right.
\end{equation}
where
$F_{i}^{N}$ is the parton structure function
in a nucleon,
$x_{n}=
1/(2r_{p}m_{p})$,
$x_{A}$ is a saturation point ($x_{A}=1/(2R_{A}m_{p})$),
$m_p$ is the mass of the proton,
$r_{p}$ is the radius of a proton and
$R_{A}$ is the
radius of the nucleus.  The only free parameter is a
constant $D$ which can be determined by
fitting the data.

Comparison of the shadowing function with
all
deep inelastic lepton-nucleus data on the
ratio $F_2^A(x,Q^2)/F_2^D(x,Q^2)$ [6]
indicate that Eq. (8) has
much steeper x-dependence, especially
for
$0.002\leq x \leq 0.1$, the region
of relevance to charm production
at RHIC and LHC energies.
In addition, the onset of saturation for the ratio of
structure functions for Xe to Deuterium is observed
at
values of $x$ about order of magnitude smaller than those
predicted by the parton recombination model [6].
Thus, it is not
surprising that even
the best fit of Eq. (8) to the data
overestimates the observed shadowing
effect by about
$15\%$.
Consequently
charm production calculated with this shadowing function
would be underestimated
by about $40\%$.

In our calculation we use
the shadowing function that has
recently been proposed as the best fit of
EMC,
NMC and E665 data [11] and is given by [13]
\begin{equation}
R(x,A)
=\left\{\begin{array}{ll}
\alpha_3 -\alpha_4 x & x_0 <x\leq 0.6 \\
(\alpha_3 -\alpha_4 x_0)\frac{1+k_q \alpha_2 ({1/x}-1/x_{0})}
{1+k_q A^{\alpha_1}({1/x}-1/x_{0})}
&x\leq x_{0}\\
\end{array}
\right.
\end{equation}
The values for the parameters $k_q$, $\alpha_{1}$,
$\alpha_2$, $\alpha_3$, $\alpha_4$ and $x_0$ can be found in Ref. 13.
It is interesting to note that
for $x \sim x_0=0.14$, this shadowing function
has a form similar to the shadowing function motivated by the
parton recombination picture.
The main difference
is the onset of
saturation, which occurs at smaller value of $x$ than $x_A$ of
Eq. (8) and much more gradually.  In addition, the A-dependence
is much weaker, namely $A^{0.1}$.

By integrating Eq. (2) and multipling by $T_{AA}(0)$ (for central
collisions) or
by $A^2/\sigma_{in}^{AA}$ (for inelastic collisions),
we obtain rapidity distribution,
transverse momentum distribution and the total cross section for the
inclusive charm quark
production in AA collisions.  We compare our results for the total
charm cross section with the low-energy hadronic and nuclear
data [14], for the energy range $20GeV\leq \sqrt s \leq 55GeV$.
We find very good agreement with all the data for the choice
of scale $Q=m_c$.
When $Q=2m_c$ is used, our cross sections
are slightly smaller than the measured
values.  Detailed comparison
of our results with low-energy measurements of differential
distributions and the total cross section for charm production
will be published separately
in Ref. 15.

Here we present our results on rapidity and
transverse momentum
distributions for the inclusive charm production in Au-Au
collisions at RHIC.
In our calculation we
use the two-loop evolved parton structure functions, MRS-S0 [5]
with
$\Lambda_5=140MeV$.
By using two other sets of structure functions, MRS-D0 and
MRS-D-- [5], we find that
theoretical uncertainty due to the choice of
the nucleon structure function is only $10\%$.  This is not
surprising because
the average $x$-value
probed with charm production at $\sqrt s=200GeV$
is about $10^{-2}$,
which is still within the range of $x$ for which
there is deep inelastic lepton-nucleon scattering data.
Another theoretical
uncertainty is the choice of the renormalization
scale.  We find about $40\%$ lower  cross sections
when we use the scale $Q=2m_c$ instead of $Q=m_c$.
Since the low-energy
data seem to be better described with the choice of
scale $Q=m_c$, we will present
most of our results with that particular
choice .
Our results for differential distributions ($d\sigma/dy$,
$d\sigma/dp_{T}$, $d\sigma/dx_{F}$ and
$d\sigma/dp_Tdy\vert_{y=0}$)
and for the total
charm cross section in heavy-ion collisions at LHC,
(including the
theoretical uncertainties), will  be included in
Ref. 15.

\par
Our result for the rapidity distribution of
inclusive charm production in central Au-Au collisions at RHIC is
presented in Fig. 1a (solid
line).  We also show the rapidity distribution when
nuclear shadowing is not included (dotted line) and
 the
leading-order results with shadowing (short-dashed line) and without
shadowing (long-dashed line).
We note the the {\it shape} of the rapidity distribution does not
seem to be affected by
the next-to-leading corrections or
by the nuclear shadowing effect.
In Fig. 1b) we show our results for
two different choices of
scale, illustrating the uncertainty due to this particular choice.
We find that, in the central rapidity region,
the number of charm
quark pairs produced per unit rapidity in central Au-Au collisions
at RHIC
is $0.6$
for the scale $Q=2m_c$
and $0.9$ for the scale $Q=m_c$.

In hadronic collisions, one usually defines the ``K-factor''
as a measure of the size of higher-order corrections.  Here we
define the {\it effective} K-factor for {\it nuclear} collisions
as a ratio of the particular distribution
to the leading-order distribution without
any nuclear effects.  In Fig. 2 we present our results for the
K-factor.
We show that in
hadronic collisions
K-factor is about $2$ (squares), while
the
{\em nuclear}  K-factor is
about
$1.4$ (circles)
in the central rapidity region ($\vert y \vert \leq 3$).
This is due to the fact that
the nuclear shadowing effect effectively suppresses production
of charm quarks by about $30\%$.

In Fig. 3 we present our results for the transverse momentum
distribution of the
charm quark produced in Au-Au collisions at RHIC (solid line).
We find
that both higher-order correlations and the nuclear effects change
the shape of
$p_{T}$
distribution.  The nuclear shadowing effect is much stronger at
low $p_T$ (about $40\%$ effect), while at $p_T=6GeV$ it reduces the
cross section by only
few percent.
The next-to-leading order corrections give a factor of $1.7$
increase at low $p_T$ and about factor of $3$ at $p_T=6GeV$.  These
two effects together result in
effective K-factor increasing from $1$ at $p_T=1GeV$ to
$3$
at $p_T=6GeV$.  At large $p_T$, where nuclear shadowing effects are
negligible, we expect K-factor to approach its hadronic value.
We find similar behavior
of the K-factor for $x_F$ distribution, namely
its strong dependence on $x_F$ [15].

By integrating differential distributions over the phase space
we obtain the total number of charm quark pairs produced.
For central (inelastic) Au-Au collisions
we get about $4$ ($1$) charm quark pairs produced per event.

To obtain the
effective A-dependence of the
total inclusive charm cross section in nuclear collisions,
defined as
$\sigma_{tot}^{charm}=A^{\beta} \sigma_{c}^{pp}$, we
use  the total charm cross section in hadronic collisions at
$\sqrt s=200GeV$, $\sigma_c^{pp}=0.18mb$.  We find that
$\beta=1.27$ for central collisions and $\beta=1.94$ for
inelastic collisions.
\par
To be able to determine the fraction of central or inelastic
events which contain at least one charm quark pair, we need to
consider the
semiclassical probability
of having at least one parton-parton collision at fixed impact
parameter,
$1 - e^{-T_{AA}(b)\sigma_c}$,
where
$e^{-T_{AA}(b)\sigma_c}$
is the probability that there is
{\it no} parton-parton scattering in Au-Au collision at
impact parameter $b$.
The fraction of events in Au-Au
collisions that  contain at least one charm quark pair is then given by

\begin{equation}
\frac {\sigma_{c}^{AA}}{\sigma_{inelastic}^{AA}}
=\frac {\int d^{2}b[1-\exp(-T_{AA}(b){\sigma_c})]}
{\int d^{2}b[1-\exp(-T_{AA}(b)\sigma_{in}^{pp})]}
\end{equation}
where $T_{AA}$ is given by Eq. (5) and
and ${\sigma_c}$ is the integrated charm cross section
{\footnote {
The total cross section for heavy-quark production in hadronic
collisions has been previously calculated using the
Eq. (11) and found to be in agreement with UA1 data
on bottom
production at $\sqrt s=630GeV$ [16].}}

\begin{equation}
\sigma_c=\int_{\frac{4m^{2}_{c}}{s}}^{1}dx_{a}\int_{\frac{4m^{2}_{c}}
{x_{a}s}}^{1}dx_{b}\sum_{i,j}^{partons}[F_{i/A}(x_{a},Q^{2})
F_{j/A}(x_{b},Q^{2})\hat{\sigma}_{i,j}(\hat{s},m_{c}^{2},Q^{2})
\end{equation}
The parton cross section $\hat\sigma_{i,j}(\hat s, m_c^2, Q^2)$
has been calculated to the order O($\alpha_s^3$) and can be written
as [17]
\begin{equation}
\hat \sigma_{i,j}(\hat s, m_c^2, Q^2)={\alpha_s^2(Q^2)\over
{m_c^2}}
f_{i,j}(\rho, {Q^2\over{m_c^2}}),
\end{equation}
where
\begin{equation}
f_{i,j}(\rho,{Q^2/{m_c^2}})=f_{i,j}^{(0)}(\rho)+4\pi\alpha_s(Q^2)
[f_{i,j}^{(1)}(\rho)+\bar f_{i,j}^{(1)}(\rho)\ln {({Q^2/{m_c^2}})}
].\end{equation}
The functions $f_{i,j}^{(0)}$,$f_{i,j}^{(1)}$,and
${\bar f}_{i,j}^{(1)}$,  are given in Ref. 17.

To determine the fraction of
all {\it central} events that contain at least one charm quark pair we
integrate Eq. (9) over the small range of impact parameter, i.e.
$0\leq b \leq 0.1fm$.  We find that about $98\%$ of central events
will contain at least
one charm quark pair.
For {\it inelastic} collisions we integrate Eq. (9)
over all
impact parameter and find this fraction
to be $38\%$.
Note that the integrated
charm cross section in Eq. (9) includes
multiple independent parton-parton scatterings which
means
multiple charm quark pair production.

To conclude,
we
have presented the complete next-to-leading order
calculation of the
differential and total inclusive cross sections
for the charm production in Au-Au
collisions.  We have shown that at RHIC energies,
both higher-order contributions and the
nuclear shadowing effect
are large and can not be
neglected. In the central region
of the rapidity distribution,
the higher-order contributions increase the cross section
by a factor of $2$, while the nuclear shadowing effect result in
additional decrease
of
about $30\%$.
These two effects together result in
{\it nuclear} K-factor of
about $1.4$.  On the other hand,
in case of
the
$p_T$
distribution, the  K-factor changes from $1$ at $p_T=1GeV$ to
$3$ at $p_T=6GeV$.
The effective A-dependence for the charm
production in the central (inelastic) collisions is
found to be
$A^{1.27}$ ($A^{1.94}$).
By properly incorporating
multiple parton scatterings we have shown that
about $98\%$ ($38\%$)
of all central (inelastic) events at RHIC will contain charm quark
pairs.
We have found that at RHIC energies the dominant subprocess for charm
production is gg fusion (about $95\%$).
Therefore, future measurements of charm production in p-p, p-A
and A-A
collisions at RHIC energies,
in addition to being a test of perturbative QCD,
could provide valuable information about presently unknown
$A$-dependence and $x$-dependence
of the nuclear shadowing in the gluon density.

We would like to emphasize that
our calculation of charm production includes
next-to-leading order corrections and
is performed in
the standard parton model with the use of
factorization theorems for
hard scatterings in pQCD.
As a consequence of the truncation of the perturbative series,
our results
{\it are} sensitive to
the choice of the
renormalization and factorization scale.
We find this uncertainly to be about
$40\%$, which is roughly an estimate for the size
of the higher-order
corrections.
We evolve the factorization scale down to
$Q=2m_c$.
The behavior of the structure functions at scales below
$2m_c$ are not very well known [9].
Furthermore,
we do not perform our calculation
with renormalization
scale below $m_c$, because
the
perturbative calculation becomes unreliable.
The nuclear shadowing effect is incorporated in a simple way,
namely by using the parametrization of
the recent
data on shadowing of the quark densities [6]
with the assumption that the
shadowing effect for gluons is the same.
The
shadowing of the gluon density,
obtained
perturbatively by solving the modified Altarelli-Parisi
equation for the structure functions [13] is
less than the observed effect for
quarks [6].
Further theoretical work along these
lines is necessary
to have a better
understanding of
parton densities in the nucleus.
Another
approach
would be to consider the role of
quantum-mechanical interference in the heavy-ion collisions [18].
The most interesting
result of this novel approach is that both the
shadowing and the antishadowing effects
are obtained without any modification of the structure functions.
We expect
future high precision measurements of deep-inelastic p-A
scatterings and
the Drell-Yan production to
be able to test both the standard parton model picture and the
quantum-mechanical interference effect providing an important
information about
the origin of the nuclear shadowing effect.

Finally, we make a remark on the possibility of detecting  a
signal for
the formation of quark-gluon plasma via enchanced charm production
in heavy-ion collisions at RHIC.  We have found that
$0.6-.9$
open charm quark pairs per unit rapidity (in the central region)
will be produced
in central Au-Au collisions via hard parton-parton scatterings.
Even though our results seem to indicate that
unrealistically high initial temperature of the quark-gluon plasma [4]
is needed to overcome this size of the background,
further theoretical and experimental work is needed in quantitative
understanding of the nuclear shadowing effect, especially of the
gluon density, before
definite conclusion can be made.

We would like to thank M. Mangano and P. Nason
for providing us with
the fortran routines for calculating double differential
distributions in hadronic collisions.
We are grateful to R. S.  Fletcher for many stimulating
discussions and
helpful comments.  We are also indebted to Jon Bagger for his
hospitality while this work was performed.
This work was supported in part through
U.S. Department of Energy Grants Nos.
DE-FG03-93ER40792 and
DE-FG02-85ER40213.
\vskip 0.2 true in
\centerline {\bf References}
\vskip 0.2true in
\noindent
[1]  For a recent review see, {\it Quark Matter '93,
Proceedings of the Tenth International
Conference on Ultrarelativistic Nucleus-Nucleus Collisions},
ed. H. A. Gustaffson,
Nucl.\ Phys.\ {\bf A566} (1994).
\vskip 0.15true in
\noindent
[2]  For example, see
B.  Muller, in {\it Particle Production in Highly Excited Matter}, eds.
H. Gutbrod and J. Rafelski (Plenum, New York, 1993).
\vskip 0.15true in
\noindent
[3]  For a recent review of the thermal photons and dileptons see,
V. Ruuskanen, in {\it Proceedings of Quark Matter '90},
Nucl. Phys. {\bf A525} (1991) 255c and on
$J/\psi$ suppression see, S.  Gavin,
in {\it Proceedings of the
Second International Conference on Physics and Astrophysics
of Quark Gluon Plasma}, Calcutta, 19-23 January 1993, in press.
\vskip 0.15true in
\noindent
[4] E. Shuryak, Phys. Rev. Lett. {\bf
68}
(1992) 3270; K. Geiger,
Phys. Rev. {\bf D48} (1993) 4129.
\vskip 0.15true in
\noindent
[5]
A. D. Martin, W. J. Sterling and R. G. Roberts, Phys. Rev.
{\bf D47} (1993) 867.
\vskip 0.15true in
\noindent
[6]  NMC Collaboration, P. Amaudruz {\it et al.},
Z. Phys. {\bf C51} (1991) 387;
E665 Collaboration, M. R. Adams {\it et al.},
Phys. Rev. Lett. {\bf {68}} (1992) 3266; EMC Collaboration,
J. Ashman
{\it et al.}, Phys. Lett. {\bf B202} (1988) 603;
M. Arnedo {et al.}, Phys. Lett. {\bf B211} (1988) 493.
\vskip 0.15true in
\noindent
[7]  For a recent review of the factorization theorems in
perturbative QCD see, J. C. Collins, D. E. Soper and G.
Sterman, in
{\it Perturbative Quantum Chromodynamics}, ed. A. H. Mueller
(World Scientific, Singapore, 1989).
\vskip 0.15 true in
\noindent
[8]  M. L. Mangano, P. Nason and G. Ridolfi, Nucl. Phys.
{\bf B373} (1992) 295.
\vskip 0.15true in
\noindent
[9]  For a recent review of the structure functions and HERA
physics see, G. Wolf, DESY preprint, DESY 94-022.
\vskip 0.15true in
\noindent
[10] K. J. Eskola, K. Kajantie and J. Lindfors,
Nucl. Phys. {\bf B323}
(1989) 37.
\vskip 0.15true in
\noindent
[11] A. Bohr and B. R. Mottelson, Nuclear Structure I
(Benjamin, New
York, 1969) pp. 160, 223.
\vskip 0.15true in
\noindent
[12] J. Qiu, Nucl. Phys. {\bf B291} (1987) 746; K. J. Eskola,
Nucl. Phys. {\bf B400} (1993)  240.
\vskip 0.15true in
\noindent
[13]  C. J. Benesh, J. Qiu and J. P Vary, Los Alamos preprint,
LA-UR-94-784.
\vskip 0.15true in
\noindent
[14]  S. Aoki {\it et al.}, Phys. Lett. {\bf B224} (1989) 441;
S. P. K. Tavernier, Rep. Prog. Phys. {\bf 50} (1987) 1439;
S. Barlag {\it et al.}, Z. Phys. {\bf C39} (1988) 451.
\vskip 0.15true in
\noindent
[15]  I. Sarcevic and P. Valerio, University of Arizona preprint
AZPH-TH/93-20.
\vskip 0.15true in
\noindent
[16]  G. Altarelli, M. Diemoz, G. Martinelli and P. Nason,
Nucl. Phys. {\bf B308} (1988) 724; I. Sarcevic, P. Carruthers and
Q. Gao, Phys. Rev. {\bf D40}, 3600 (1989).
\vskip 0.15true in
\noindent
[17]  P. Nason, S. Dawson
and R. K. Ellis, Nucl. Phys. {\bf B329} (1989) 49; ibid.
Nucl. Phys. {\bf B303} (1988) 607.
\vskip 0.15true in
\noindent
[18]  D. Kharzeev and H. Satz, Phys. Lett, {\bf B327}, 361 (1994);
A. Bialas and Czyz,
Phys. Lett. {\bf B328}, 172 (1994).
\vfil\eject
\begin{center}
{\bf Figure Captions}
\end{center}

Fig. 1. a)  Rapidity distribution
of inclusive charm quark
production in
Au-Au collisions at RHIC,
calculated to next-to-leading order
(LO+NLO)
including nuclear shadowing (NS)
(solid line), without NS
(dotted line),
only leading-order (LO) without NS
(long-dashed line), and with NS
(short-dashed line), b)
The same calculation as above
but for
two different choices of the scale,
$Q^{2}=m_{c}^{2}$ (solid line)
and $Q^{2}=4m_{c}^{2}$ (dotted line).
\vskip 0.15true in
Fig. 2.   The
{\it effective} K-factor,
$K\equiv ({d\sigma_c\over dy})/
({d\sigma_c\over dy})_
{{LO}}$, for the distributions
presented
in Fig. 1.
The K-factor for
the next-to-leading order (LO+NLO) distribution
without
NS (squares), including NS
(circles) and for the leading-order (LO) distribution
with NS (diamonds).
Note that $K\approx 2$
for hadronic collisions (squares).
\vskip 0.15true in
Fig. 3.  The same as in Fig. 1a)  but for the transverse
momentum distribution of inclusive charm quark production.
Curves are labeled as in Fig. 1a).
\vskip 0.15true in
Fig. 4.
The {\it effective} K-factor for the transverse momentum
distributions presented in Fig. 3.
Labeling is the same as in Fig.
2.
\end{document}